\documentclass[aps,pra,twocolumn,amsmath,amssymb,superscriptaddress]{revtex4-2}
\usepackage[english]{babel}
\usepackage{xcolor}
\usepackage{amssymb}
\usepackage{dcolumn}
\usepackage{bm}

\usepackage{amsmath}
\usepackage{braket}

\allowdisplaybreaks[2]   %Formula span

\usepackage{graphicx}        % standard LaTeX graphics tool

\graphicspath{{pict/}{}}

\usepackage[normalem]{ulem}

\usepackage[pdfstartview=FitH, CJKbookmarks=true, bookmarksnumbered=true, bookmarksopen=true, colorlinks=true, pdfborder=001, citecolor=blue, linkcolor=blue, urlcolor=blue, linktocpage=true] {hyperref}

\begin{document}

\title{Spontaneous Chiral Superradiance of a Spin-\(1\) Condensate in a Ring Cavity}

\author{Jingjun You}
\thanks{These authors contributed equally to this work.}
\affiliation{Guangdong Provincial Key Laboratory of Quantum Metrology and Sensing $\&$ School of Physics and Astronomy, Sun Yat-Sen University, Zhuhai 519082, China}

\author{Haoran Jia}
\thanks{These authors contributed equally to this work.}
\affiliation{Guangdong Provincial Key Laboratory of Quantum Metrology and Sensing $\&$ School of Physics and Astronomy, Sun Yat-Sen University, Zhuhai 519082, China}

\author{Xinyu Feng}
\affiliation{Guangdong Provincial Key Laboratory of Quantum Metrology and Sensing $\&$ School of Physics and Astronomy, Sun Yat-Sen University, Zhuhai 519082, China}

\author{Zhenhua Yu}
\email{yuzhh5@mail.sysu.edu.cn}
\affiliation{Guangdong Provincial Key Laboratory of Quantum Metrology and Sensing $\&$ School of Physics and Astronomy, Sun Yat-Sen University, Zhuhai 519082, China}

\author{Su Yi}
\email{yisu@nbu.edu.cn}
\affiliation{Institute of Fundamental Physics and Quantum Technology $\&$ School of Physical Science and Technology, Ningbo University, Ningbo, 315211, China}
\affiliation{Zhejiang Key Laboratory of Extreme Universe, Ningbo University, Ningbo, Zhejiang 315211, China}

\author{Yuangang Deng}
\email{dengyg3@mail.sysu.edu.cn}
\affiliation{Guangdong Provincial Key Laboratory of Quantum Metrology and Sensing $\&$ School of Physics and Astronomy, Sun Yat-Sen University, Zhuhai 519082, China}

\date{\today}

\begin{abstract}
We investigate superradiant phase transitions in a spin-$1$ Bose-Einstein condensate coupled to a ring cavity, where two symmetry-related Raman processes couple the atomic spin states to counterpropagating cavity modes. Above the superradiant threshold, we find that the symmetry between the two Raman channels is spontaneously broken, giving rise to a chiral superradiant (CSR) phase. Despite the exchange symmetry of the Hamiltonian, the CSR state spontaneously selects one of the two propagation directions, with one cavity mode becoming macroscopically occupied while its counterpropagating partner remains dark. Spin-dependent collisions further reshape the superradiant phase diagram by selectively shifting the boundaries between the chiral and symmetric superradiant phases, a feature intrinsic to the multicomponent spin-$1$ manifold. For sufficiently negative spin-dependent interactions, the interaction-induced modification of the atomic level structure gives rise to a reversed plane-wave phase characterized by an inverted internal-state occupation. These results identify spontaneous selection between symmetry-related superradiant channels as a mechanism for generating chirality in many-body cavity QED and demonstrate how spin-dependent collisions control the resulting superradiant phase structure.
\end{abstract}

\maketitle

\section{Introduction}

Phase transitions provide a general framework for understanding the emergence of collective behavior in many-body systems~\cite{RevModPhys.46.437,RevModPhys.68.473,RevModPhys.69.315,
RevModPhys.80.167,RevModPhys.82.2155,RevModPhys.89.040501}. A paradigmatic example in cavity QED is the superradiant phase
transition, where sufficiently strong collective light-matter coupling gives rise to a macroscopically occupied cavity field~\cite{PhysRev.93.99,HEPP1973360,PhysRevA.7.831}. Such transitions have been realized with Bose-Einstein condensates~\cite{baumann2010dicke,mottl2012roton,leonard_supersolid_2017,leonard_monitoring_2018,PhysRevLett.118.013603}
and degenerate Fermi gases~\cite{doi:10.1126/science.abd4385,PhysRevLett.112.143002,PhysRevLett.112.143003,PhysRevLett.112.143004}. Engineering interactions~\cite{z8gv-7yyk,PhysRevLett.134.193603} and spatial configurations~\cite{qj5x-t71k,rbs2-2pd5} further broadens the scope of superradiance, giving access to collective phenomena such as superradiant neutrino emission~\cite{l3c1-yg2l} and collective superradiant decay
~\cite{w377-f9mx}.

Ultracold atomic gases coupled to optical cavities provide a versatile platform for exploring quantum many-body physics~\cite{RevModPhys.80.885,RevModPhys.83.863,RevModPhys.83.1523,RevModPhys.91.015005,gross2017quantum} and cavity-enabled quantum technologies~\cite{RevModPhys.87.1379,RevModPhys.94.041003,RevModPhys.93.025005}. The tunability of light-matter coupling and cavity-mediated interactions enables control over both internal and motional atomic degrees of freedom, giving access to a broad range of collective
phases and dynamical phenomena~\cite{PhysRevLett.128.103201,RevModPhys.85.553,Mivehvar:2021lpo,RevModPhys.91.015006,RevModPhys.85.299,RevModPhys.95.035002}. Cavity-induced superradiance in ultracold gases has been studied predominantly in spinless~\cite{baumann2010dicke,mottl2012roton,leonard_supersolid_2017,leonard_monitoring_2018,PhysRevLett.118.013603} and spin-$1/2$ systems~\cite{You:25,PhysRevLett.122.193605,PhysRevLett.121.163601}, where collective atom-light coupling gives rise to Dicke- or Tavis-Cummings-type transitions and, in spinful settings, dynamical spin-orbit coupling~\cite{PhysRevLett.123.160404,PhysRevLett.120.123601,PhysRevLett.112.143007,PhysRevA.89.011602,PhysRevResearch.5.013002}.

Ring cavities introduce an additional channel degree of freedom through the coupling of atoms to two counterpropagating optical modes~\cite{PhysRevLett.122.190801,PhysRevLett.113.073003,PhysRevLett.124.033601}. In conventional ring-cavity superradiance, the two symmetry-related modes can become simultaneously occupied, preserving the channel-exchange symmetry. This raises a central question: must superradiance preserve the channel-exchange symmetry, or can the many-body system spontaneously select only one of two symmetry-related emission channels? Directional optical responses are commonly achieved by explicitly breaking the symmetry between opposite propagation directions~\cite{PhysRevLett.128.083604,PhysRevLett.131.113602,PhysRevLett.132.193602}, for example by breaking time-reversal~\cite{PhysRevLett.132.210402} or spatial-inversion symmetry~\cite{ando_observation_2020}. In cavity QED, such asymmetries can be introduced through cavity rotation or unidirectional pumping to realize directional superradiance and quantum squeezing~\cite{PhysRevLett.132.193602,PhysRevLett.128.083604}. More generally, explicit symmetry breaking provides a common route to directional transport and asymmetric transmission in nonreciprocal photonic systems~\cite{PhysRevLett.121.153601,PhysRevLett.106.213901,doi:10.1126/science.1214383,PhysRevLett.102.213903,PhysRevLett.118.033901,PhysRevLett.110.093901,PhysRevLett.121.203602,PhysRevLett.120.043901}.

A qualitatively different scenario arises when the Hamiltonian remains symmetric under channel exchange, while the many-body state spontaneously selects one of the two symmetry-related configurations. Such spontaneous selection provides a route to emergent chirality~\cite{zhang_interface_2021,del_pino_non-hermitian_2022}, a concept explored in systems ranging from chiral molecules and magnons~\cite{gn6c-1q19} to plasmonic metasurfaces~\cite{kf83-1b8s}. Recent developments in magnetic, photonic, and quantum-optical systems have further highlighted the role of symmetry breaking in generating and controlling chiral responses~\cite{m8lc-f8gk,nature654.926,nature651.341,doi:10.1126/science.ado1611,8qx2-xxh2,doi:10.1126/science.adx5963,nature654.369}. In many realizations, the imbalance between opposite channels is imposed by external fields, geometric asymmetry, or intrinsic material handedness. By contrast, whether a superradiant transition can spontaneously select one of two symmetry-related channels and thereby generate chirality without an explicit directional bias remains largely unexplored.

This question acquires an additional dimension in higher-spin condensates, where collisional interactions can couple channel selection to the internal-state energetics. In commonly studied spinless and spin-$1/2$ cavity systems, contact interactions often provide only weak corrections to the superradiant threshold~\cite{baumann2010dicke,mottl2012roton,leonard_supersolid_2017,leonard_monitoring_2018,PhysRevLett.123.160404,PhysRevLett.121.163601,PhysRevLett.120.123601}. Spin-$1$ condensates, by contrast, possess additional spin-dependent collisional channels~\cite{PhysRevLett.81.5257} that can modify the relative energetics of competing internal-state configurations. These interactions can therefore modify not only the superradiant phase boundaries but also the internal-state configuration underlying the collective Raman processes.

Here we show that a superradiant transition can spontaneously select between two symmetry-related Raman channels in a spin-$1$ Bose-Einstein condensate coupled to a ring cavity. Above the superradiant threshold, the channel-exchange symmetry of the Hamiltonian is spontaneously broken, yielding a chiral superradiant (CSR) phase in which one counterpropagating cavity mode becomes macroscopically occupied while the other remains dark. The resulting CSR phase originates from spontaneous competition between the two symmetry-related superradiant channels, rather than
from an externally imposed directional bias. Spin-dependent collisions play a complementary role by modifying the relative energetics of the chiral and symmetric superradiant states and thereby shifting their phase boundaries. For sufficiently negative spin-dependent interactions, the interaction-induced modification of the internal-state energetics further gives rise to a reversed plane-wave phase with inverted atomic-state occupation. Our results establish spontaneous selection between symmetry-related superradiant channels as a route to emergent chirality in cavity QED and show how spin-dependent interactions reshape the associated many-body phase structure~\cite{bchn-b47c,8qx2-xxh2,doi:10.1126/science.adx5963,nature654.369}.

The remainder of this paper is organized as follows. In Sec.~II, we introduce the spin-$1$ atom--ring-cavity model and derive the corresponding effective Hamiltonian. In Sec.~III, we present the phase diagram and characterize the chiral and plane-wave superradiant phases. In Sec.~IV, we summarize our main results and discuss their experimental implications.

\section{Model}
%\subsection{Microscopic Hamiltonian}

\begin{figure}
\includegraphics[width=\columnwidth]{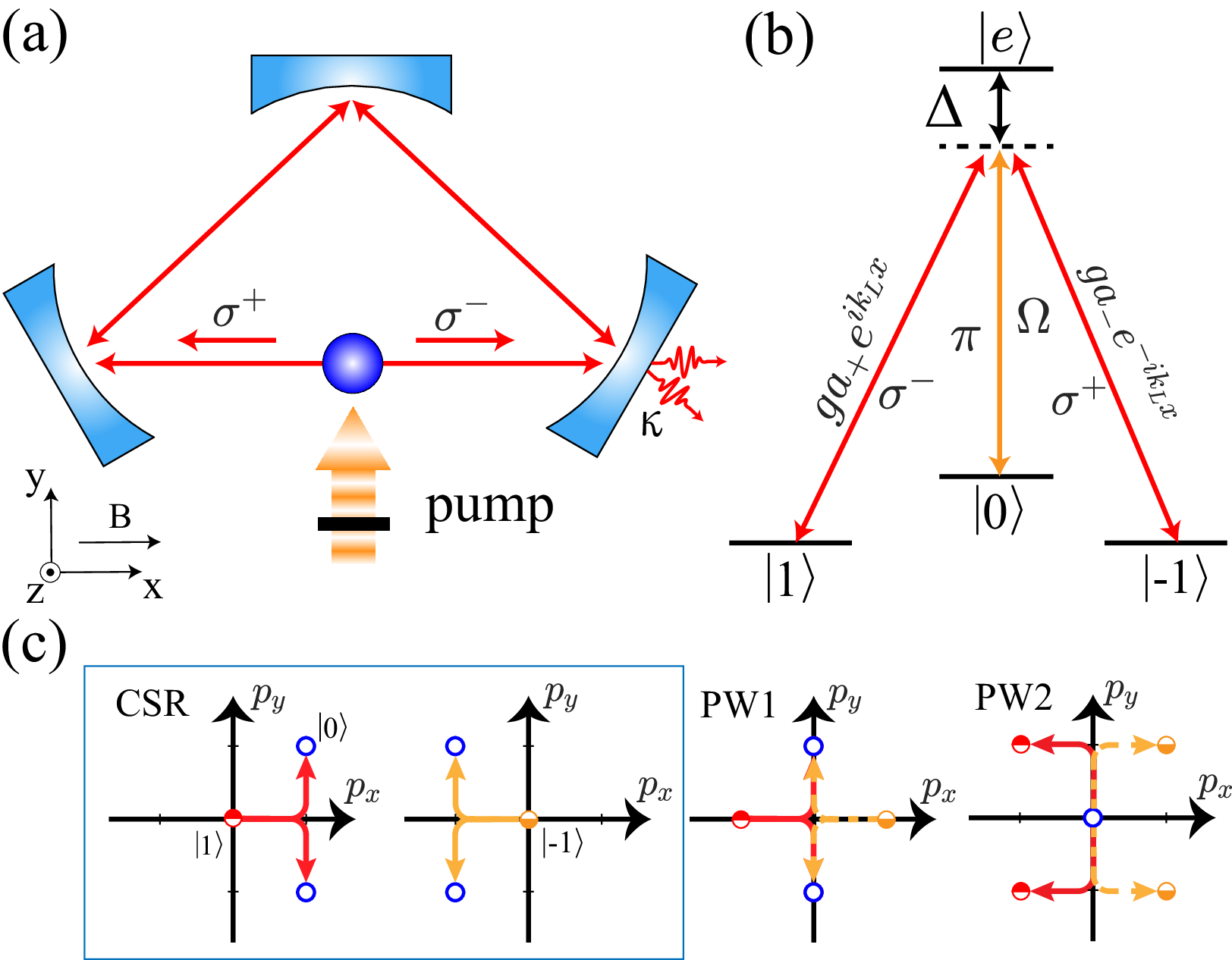}%
\caption{(a) Schematic of a spin-$1$ BEC in a ring cavity. (b) Atomic level scheme. (c) Momentum-space Raman pathways for the two symmetry-related CSR states and the PW1 and PW2 phases (from left to right). Blue open, orange lower-half-filled, and red upper-half-filled circles denote the $m_F=0,-1$, and $+1$ components, respectively; orange and red arrows indicate the two Raman channels.}
\label{model}
\end{figure}

We consider a Bose-Einstein condensate of $N$ spin-1 atoms coupled to two counterpropagating modes of a ring cavity, as illustrated in Fig.~\ref{model}(a). An external magnetic field $\mathbf B$ defines the quantization axis and lifts the degeneracy of the spin states $|m_F=0,\pm1\rangle$, as shown in Fig.~\ref{model}(b). We neglect the linear Zeeman shift while retaining the quadratic Zeeman shift $\hbar\delta$~\cite{PhysRevLett.132.093402}. An off-resonant standing-wave pump propagating along the $y$ direction couples $|0\rangle$ to the excited state $|e\rangle$ with Rabi frequency $\Omega$. The two counterpropagating cavity modes $\hat a_\pm$ couple $|\pm1\rangle$ to $|e\rangle$, respectively, with spatially dependent single-atom couplings $g_\pm(x)=g e^{\pm ik_Lx}$, where $k_L=2\pi/\lambda$ is the wave vector of the laser field and $\lambda$ is its wavelength. Together with the $\pi$-polarized pump, the two $\sigma$-polarized cavity fields generate two symmetry-related Raman processes that couple the atomic spin and motional degrees of freedom.

In the dispersive regime $|\Delta|\gg g,\Omega$, the electronically excited state $|e\rangle$ can be adiabatically eliminated. The resulting many-body Hamiltonian for the spin-1 condensate and the two cavity modes is
\begin{align}
\hat H
&=
\hat H_{\rm sp}+\hat H_{\rm int},
\\
\hat H_{\rm sp}
&=
\int d^3\mathbf r\,
\hat{\boldsymbol{\Psi}}^\dagger(\mathbf r)
\left[
-\frac{\hbar^2\nabla^2}{2M}
+V_T(\mathbf r)
+\hbar\hat{\boldsymbol h}(\mathbf r)
\right]
\hat{\boldsymbol{\Psi}}(\mathbf r)
\nonumber\\
&\quad
+\hbar\Delta_c
\sum_{\nu=\pm}
\hat a_\nu^\dagger\hat a_\nu ,
\label{Hsp}
\\
\hat H_{\rm int}
&=
\int d^3\mathbf r
\left[
\frac{g_s}{2}\hat n^2
+\frac{g_a}{2}\hat S_z^2
+\frac{g_b}{2}\hat n_0^2
\right].
\label{Hint}
\end{align}
Here $M$ is the atomic mass, $\hat{\boldsymbol{\Psi}}=(\hat\psi_1,\hat\psi_0,\hat\psi_{-1})^T$ is the spinor field operator, and $\hat a_\pm$ annihilates a photon in the two counterpropagating cavity modes. We define the component densities $\hat n_m=\hat\psi_m^\dagger\hat\psi_m$, the total density $\hat n=\sum_m\hat n_m$, and the longitudinal spin density $\hat S_z=\hat n_1-\hat n_{-1}$. The short-range interaction strengths $g_s$, $g_a$, and $g_b$ are given by $g_j=4\pi\hbar^2a_j/M$, where $a_j$ is the corresponding $s$-wave scattering length in the interaction channel $j=s,a,b$. In the following, $g_b$ serves as the control parameter for the interaction-driven reconstruction of the phase diagram~\cite{PhysRevA.108.023308,chd2-slgy,PhysRevA.79.023401,Helson2023}. In the basis $(|1\rangle,|0\rangle,|-1\rangle)$, the single-particle Hamiltonian entering Eq.~(\ref{Hsp}) is given by
\begin{align}
    \hat{{\boldsymbol h}}=
    \begin{pmatrix}
        M_0(y) &M_+^\dagger &U_0\hat{a}_+^\dagger\hat{a}_-e^{-2ik_Lx}\\
        M_+&M_0(y)+\delta&M_-\\
        U_0\hat{a}_-^\dagger\hat{a}_+e^{2ik_Lx}&M_-^\dagger& M_0(y)
    \end{pmatrix}\label{single}
\end{align}
where $M_0(y) = U_0(\hat{a}^\dagger_+\hat{a}_+ +\hat{a}^\dagger_-\hat{a}_-) +U_y\cos^2(k_L y)$ contains the ac Stark shifts induced by the cavity and pump fields, with $U_0 = -g^2/\Delta$ and $U_y=-\Omega^2/\Delta$. The effective Raman couplings are $M_\pm=\Omega_0\cos(k_L y)e^{\pm ik_L x}\hat{a}_\pm$ with $\Omega_0=-g\Omega/\Delta$ being the maximum effective Raman coupling strength.

Importantly, the two Raman processes are related by a discrete channel-exchange symmetry. In the absence of an explicit directional bias, the Hamiltonian is invariant under the simultaneous transformation
\[
\hat a_+\leftrightarrow\hat a_-,
\qquad
\hat\psi_1\leftrightarrow\hat\psi_{-1},
\qquad
x\rightarrow -x.
\]
Thus, neither propagation direction is favored at the Hamiltonian level. As shown below, the CSR phase spontaneously breaks this symmetry by selecting only one of the two Raman-cavity channels.

In the large-detuning regime $|\widetilde{\Delta}_c|\gg\{\Omega_0,|U_0|,|U_y|\}$, the cavity dynamics occurs on a much faster time scale than the
external atomic motion. The cavity fields therefore adiabatically follow the atomic configuration and can be eliminated by setting their equations of
motion to the stationary limit. Introducing the coherent cavity amplitudes $\alpha_\pm=\langle\hat a_\pm\rangle$, we obtain
\begin{align}
\alpha_+
&=
\frac{
U_0\Omega_0\mathcal G\,\Xi_{-1}
-\Omega_0(\widetilde{\Delta}_c-i\kappa)\Xi_1
}{
\mathcal D_c
},
\nonumber\\
\alpha_-
&=
\frac{
U_0\Omega_0\mathcal G^*\,\Xi_1
-\Omega_0(\widetilde{\Delta}_c-i\kappa)\Xi_{-1}
}{
\mathcal D_c
},
\label{cavity_ss}
\end{align}
where $\mathcal D_c=
(\widetilde{\Delta}_c-i\kappa)^2
-U_0^2|\mathcal G|^2 
$, $\mathcal G
=
\langle\psi_1|
e^{-2ik_Lx}
|\psi_{-1}\rangle$, and $\Xi_{\pm1}
=
\langle\psi_{\pm1}|
\cos(k_Ly)e^{\mp ik_Lx}
|\psi_0\rangle$. Here $\mathcal G$ characterizes the coherence between the $m_F=\pm1$ components, while $\Xi_{\pm1}$ quantify the two Raman
scattering channels connecting $m_F=0$ to $m_F=\pm1$. Because the two counterpropagating modes belong to the same cavity and have the same decay rate $\kappa$, they evolve on the same fast optical time scale and can be adiabatically eliminated simultaneously. The resulting amplitudes $\alpha_\pm$ depend self-consistently on the atomic wave functions through $\mathcal G$ and $\Xi_{\pm1}$, providing the nonlinear feedback that underlies the competition between the two Raman-cavity channels.

To gain insight into the microscopic origin of the CSR phase, we project the atomic field operators onto the dominant momentum modes involved in the Raman processes. Retaining only the lowest momentum states connected by single-photon recoil and neglecting two-body interactions, as in the corresponding spin-$1/2$ treatment~\cite{PhysRevResearch.5.013002,You:25}, we obtain the following low-energy Hamiltonian:
\begin{align}
\hat H_{\rm CSR}/\hbar
={}&
\omega_0
\left(
\hat J_z^{(+)}+\hat J_z^{(-)}
\right)
+
\widetilde{\Delta}_c
\sum_{\nu=\pm}
\hat a_\nu^\dagger\hat a_\nu
\nonumber\\
&+
\frac{\Omega_0}{\sqrt{2}}
\sum_{\nu=\pm}
\left(
\hat a_\nu\hat J_+^{(\nu)}
+
\hat a_\nu^\dagger\hat J_-^{(\nu)}
\right),
\label{HCSR}
\end{align}
where $\hat{J}_z^{(\pm)}=(\hat{c}^\dagger_{0,\bf{k}_\pm}\hat{c}_{0,\bf{k}_\pm}-\hat{c}^\dagger_{\pm,\bf{0}}\hat{c}_{\pm,\bf{0}})/2$, $\hat{J}_-^{(\pm)}=\hat{c}^\dagger_{\pm,\bf{0}}\hat{c}_{0,\bf{k}_\pm}$ are collective spin operators and $\omega_0 = 2E_L/\hbar + \delta$ is the effective detuning of atomic field. Here $\hat{c}_{\sigma,\mathbf{k}}$ denotes the annihilation operator for an atom in spin state $\sigma=\pm1,0$ and momentum $\mathbf{k}$. For the CSR phase, $\hat c_{\nu,\mathbf 0}$ annihilates an atom in the spin state $|\nu1\rangle$ with zero momentum, whereas $\hat c_{0,\mathbf k_\nu}$ denotes the recoil mode of the $m_F=0$ component associated with Raman channel $\nu=\pm$. The two channels involve opposite momentum transfer along the cavity axis, as illustrated in the first two panels of Fig.~\ref{model}(c). The channel index $\nu=\pm$ labels the two symmetry-related Raman processes coupled to the counterpropagating cavity modes $\hat a_\nu$. As illustrated in the first two panels of Fig.~\ref{model}(c), the two channels are related by inversion of the momentum transfer along the cavity axis. Although this reduced model captures the basic two-channel structure of the superradiant coupling, it does not quantitatively reproduce the interaction-dependent phase boundaries obtained from the full many-body model. Applying the same momentum-mode projection to the PW2 phase yields
\begin{align}
\hat H_{\rm PW2}/\hbar
={}&
\omega_0\hat K_z
+
\widetilde{\Delta}_c
\sum_{\nu=\pm}
\hat a_\nu^\dagger\hat a_\nu
\nonumber\\
&+
\frac{\Omega_0}{\sqrt{2}}
\left(
\hat a_+\hat K_-^{(+)}
+
\hat a_-\hat K_-^{(-)}
+\mathrm{H.c.}
\right)
\nonumber\\
&+
U_0
\left(
\hat a_+^\dagger\hat a_-
\hat d_+^\dagger\hat d_-
+
\hat a_-^\dagger\hat a_+
\hat d_-^\dagger\hat d_+
\right),
\label{HPW2}
\end{align}
where $\hat K_z
=
\frac{1}{2}
\left(
\hat d_+^\dagger\hat d_+
+
\hat d_-^\dagger\hat d_-
-
\hat d_0^\dagger\hat d_0
\right)$ and $\hat K_-^{(\pm)}
=
\hat d_0^\dagger\hat d_\pm$ are collective spin operators. Here $\hat d_0$ annihilates an atom in the $m_F=0$ zero-momentum mode, whereas $\hat d_+\) and \(\hat d_-$ annihilate atoms in the $m_F=+1$ and $m_F=-1$ recoil modes, respectively, as illustrated in the last panel of Fig.~\ref{model}(c). The corresponding recoil momenta are $(-k_L,\pm k_L)$ and $(k_L,\pm k_L)$, respectively.

Equation~(\ref{HCSR}) has the structure of two symmetry-related Tavis-Cummings channels with identical coupling strengths. At the Hamiltonian level, the two channels are therefore exactly degenerate, and neither propagation direction is energetically preferred. The CSR phase nevertheless develops a finite cavity field in only one channel, while the symmetry-related counterpropagating mode remains dark. Because the full Hamiltonian, including the interaction terms, remains invariant under channel exchange, the two single-channel configurations are degenerate and are mapped onto each other by the discrete symmetry. The directional cavity occupation therefore does not originate from an explicit bias of the Hamiltonian, but from spontaneous breaking of the channel-exchange symmetry. The two possible CSR configurations correspond to the degenerate broken-symmetry sectors $(\alpha_+,\alpha_-)\sim(\alpha,0)$ and $(0,\alpha)$, respectively. The reduced model thus exposes the symmetry structure of the two competing superradiant channels, while the interaction terms omitted in Eq.~(\ref{HCSR}) determine their nonlinear competition and the stability of the CSR state in the full many-body problem.

\begin{figure}
\centering
\includegraphics[width=\columnwidth]{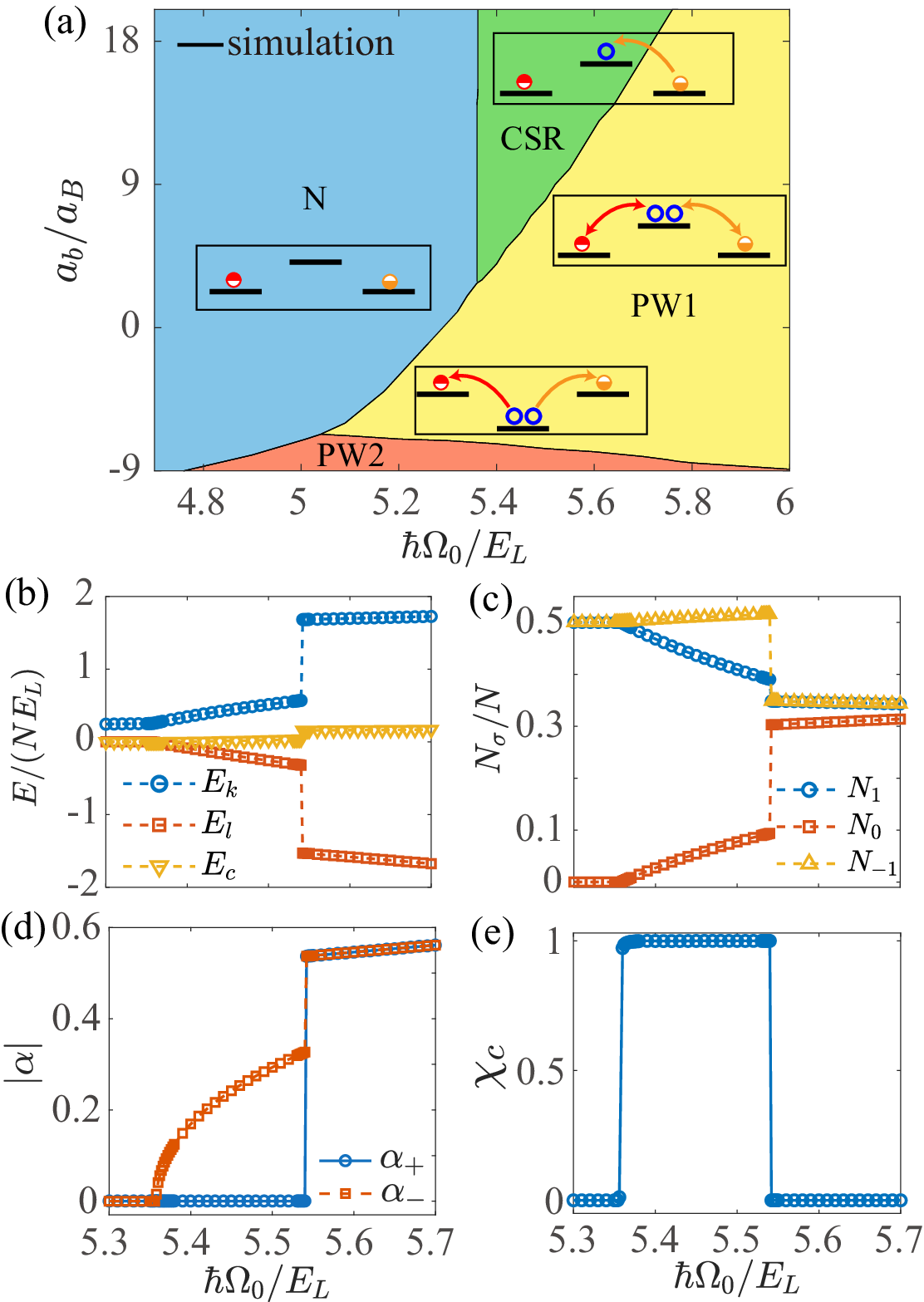}%
\caption{(a) Phase diagram in the $(\Omega_0,a_b)$ plane. Solid lines mark the numerical phase boundaries. (b) Energy, (c) spin populations, (d) cavity-field amplitudes, and (e) chiral order parameter $\chi_c$ versus $\Omega_0$ for $a_b=10a_B$. The other parameters are $U_0=-E_L/\hbar$, $\delta=1.6E_L/\hbar$, and $\widetilde{\Delta}_c=3NE_L/\hbar$. }
\label{pd}
\end{figure}

\section{Results and Discussion}

We determine the stationary mean-field states of the condensate-cavity system by solving the coupled Gross-Pitaevskii equations, with the steady-state cavity amplitudes $\alpha_\pm$ evaluated self-consistently from the atomic wave functions. The atomic mean-field configurations are obtained by imaginary-time evolution, which minimizes the corresponding mean-field energy functional $\mathcal{F}[\psi_\sigma]=\langle \mathcal{H}\rangle$ under the self-consistency condition for $\alpha_\pm$. We consider a condensate of $N=10^5$ atoms confined in a harmonic trap. In our numerical simulation, we take single-photon recoil energy $E_L/\hbar = 2\pi \times 3.53\,\text{kHz}$ corresponding to the wavelength $\lambda=2\pi/k_L=803.2\, \text{nm}$, cavity decay rate $\kappa= 100E_L/\hbar$, two-photon detuning $\delta = 1.6E_L/\hbar$, Stark shift $U_0=-1E_L/\hbar$, and pump-cavity detuning $\tilde{\Delta}_c= 3NE_L/\hbar$. The interaction strengths $g_s$ and $g_a$ correspond to the $s$-wave scattering lengths $a_s=100.4a_B$ and $a_a=10a_B$, respectively~\cite{deng2021universalrelationsatomicdipolar,PhysRevLett.88.093201}, where $a_B$ is the Bohr radius. We therefore use the Raman coupling $\Omega_0$ and the spin-dependent interaction $g_b$ as the principal control parameters.

\begin{figure}
\includegraphics[width=\columnwidth]{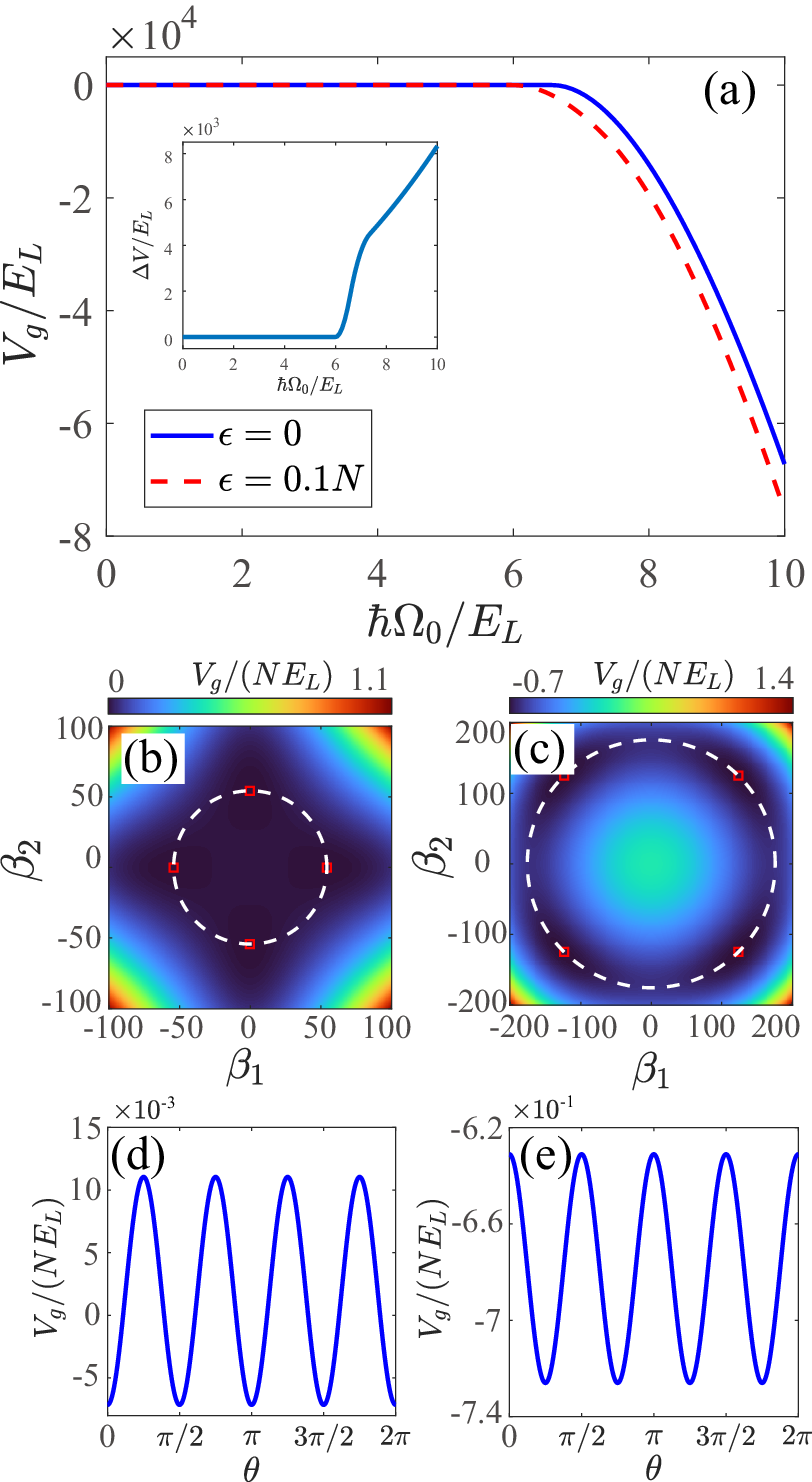}%
\caption{(a) Minimum effective potential versus the Raman coupling strength $\Omega_0$ for $\epsilon=0$ and $0.1N$; the inset shows the corresponding energy difference. (b) Effective potential $V_g(\beta_1,\beta_2)$ in the CSR regime for $\hbar\Omega_0/E_L=7$ with $a_b=3.44a_B$. (c) Effective potential $V_g(\beta_1,\beta_2)$ in the PW2 regime for $\hbar\Omega_0/E_L=7$ and $a_b=-10a_B$. Open red squares mark the four degenerate minima in (b) and (c). (d,e) Angular cuts of the potentials in (b,c), respectively, along the dashed circles passing through the minima, parametrized by the polar angle $\theta$.}
\label{pot}
\end{figure}

Figure~\ref{pd}(a) shows the mean-field phase diagram that characterizes the superradiant phase transition within the condensate-cavity system in the $(\Omega_0,a_b)$ plane. Above the superradiant threshold, three distinct superradiant phases appear: CSR, PW1, and PW2. The PW2 phase is distinguished from PW1 by an inversion of the internal-state occupation: its zero-momentum component predominantly occupies $m_F=0$, rather than $m_F=\pm1$. Correspondingly, the dominant Raman process is reversed relative to PW1, transferring atoms from $m_F=0$ to $m_F=\pm1$, as illustrated in Fig.~\ref{model}(c). 

A notable feature of Fig.~\ref{pd}(a) is the strongly phase-dependent influence of $a_b$: while the N-to-CSR boundary is only weakly affected, the boundaries involving the PW phases shift substantially with the interaction strength. This behavior contrasts with typical spin-$1/2$ condensate-cavity systems, where contact interactions produce only weak corrections to the superradiant threshold~\cite{leonard_supersolid_2017,leonard_monitoring_2018}. The interaction dependence originates from the different collisional energy shifts of the spin components, which modify the effective energy separation between the Raman-coupled atomic states. In the corresponding spin-$1/2$ setting, the dominant contact interaction shifts the relevant atomic configurations nearly equally and therefore produces only a weak change in the effective Raman detuning. Consequently, the superradiant threshold can often be understood to leading order from the atom-light coupling alone. For spin-$1$ atoms, by contrast, the additional spin-dependent interaction shifts the competing internal-state configurations differently, thereby modifying their relative energetics and reconstructing the superradiant phase boundaries.

To quantify the interaction dependence of the phase boundaries, we compare the numerical results with the critical coupling predicted by the effective Tavis-Cummings model~\cite{PhysRevResearch.5.013002,You:25}. For the PW1 boundary, however, the single-recoil truncation underlying the reduced model is no longer quantitatively accurate, and the interaction-induced shifts of the atomic modes must be retained. A closed analytical expression for the boundary is therefore not available within this approximation. We therefore use the Tavis-Cummings critical condition as a phenomenological guide, incorporating the effect of $a_b$ through an interaction-dependent effective atomic detuning. In contrast, the N-to-CSR boundary depends only weakly on $a_b$, consistent with the predominantly atom-light origin of the superradiant instability in this regime. Quantitative deviations from the reduced model nevertheless remain because the full spin-$1$ system contains additional momentum modes and interaction channels absent from the two-level approximation.

The interaction dependence of the phase boundaries can be understood qualitatively in terms of an interaction-renormalized two-photon detuning $\widetilde{\delta}$. As $a_b$ increases, the critical Raman couplings associated with the PW1 and PW2 boundaries shift to larger values of $\Omega_0$. Within this picture, increasing $a_b$ increases $\widetilde{\delta}$, enlarging the effective energy separation between the Raman-coupled atomic configurations and thereby requiring a stronger Raman coupling to reach the corresponding superradiant phase. For sufficiently negative $a_b$, however, $\widetilde{\delta}$ decreases and can change sign, corresponding to an inversion of the relative energetics of the Raman-coupled atomic configurations. This interaction-induced inversion provides the microscopic origin of the reversed internal-state occupation that distinguishes PW2 from PW1. Upon further increasing $\Omega_0$, the redistribution of the atomic populations restores the internal-state configuration characteristic of PW1, resulting in the PW2-to-PW1 transition.

In contrast, the N-to-CSR boundary is only weakly dependent on $a_b$. Near this boundary, the instability predominantly involves a single Raman channel and the two atomic modes connected by it, while the remaining spin component plays only a minor role. The transition therefore approaches an effective two-level description, for which the interaction-induced correction to $\widetilde{\delta}$ is weak.

To characterize the nature of these transitions, Figs.~\ref{pd}(b)-\ref{pd}(e) show the energy, atomic-state populations, intracavity photon numbers, and cavity-mode imbalance as functions of $\Omega_0$. At $\hbar\Omega_0/E_L\simeq5.37$, the $m_F=0$ population and the photon number of one cavity mode emerge continuously from zero, together with the onset of a finite light-matter contribution to the energy. This behavior identifies the N-to-CSR transition as continuous. At $\hbar\Omega_0/E_L\simeq5.54$, by contrast, the atomic populations and cavity occupations exhibit discontinuous jumps, accompanied by a nonanalytic change in the energy, indicating a first-order transition from the CSR phase to the PW1 phase.

Close to threshold, the system lowers its energy by activating only one of the two degenerate Raman channels, producing the intermediate CSR phase before the symmetric PW1 state becomes favorable at stronger coupling. The microscopic origin of this channel selection is analyzed below in terms of the effective energy landscape. In the CSR phase, the dominant $m_F=\pm1$ component remains at zero momentum, while only the recoil mode associated with the selected Raman channel acquires a finite population. The two CSR configurations are related by the channel-exchange symmetry: interchanging the $m_F=+1$ and $m_F=-1$ components simultaneously exchanges the two counterpropagating cavity modes. Consequently, only one of the two symmetry-related Raman channels becomes macroscopically active, providing the microscopic signature of the CSR phase.

Unlike directional responses generated by an explicit asymmetry between the two propagation directions, the present chirality emerges spontaneously from a Hamiltonian that remains invariant under channel exchange. The weak variation of the nominally inactive $m_F=-1$ population with $\Omega_0$ results from residual interaction-induced population redistribution and remains much smaller than the population change in the active Raman channel. As shown in Fig.~\ref{pd}(d), the two counterpropagating cavity modes are symmetry related and therefore equivalent at the Hamiltonian level. In the CSR phase, one mode acquires a macroscopic occupation while the other remains dark. The two possible solutions, $(\alpha_+,\alpha_-)\sim(\alpha,0)$ and $(0,\alpha)$, are degenerate and related by channel exchange. Upon increasing $\Omega_0$ further, the system enters PW1, where the channel-exchange symmetry is restored and the two cavity modes acquire equal finite occupations.

To characterize the spontaneous breaking of the channel-exchange
symmetry, we introduce the signed chiral order parameter
\begin{equation}
\chi_c=\frac{|\alpha_-|^2-|\alpha_+|^2} {|\alpha_+|^2+|\alpha_-|^2},
\label{chiral_order}
\end{equation}
Under channel exchange $\alpha_+\leftrightarrow\alpha_-$, $\chi_c\rightarrow-\chi_c$, such that $\chi_c$ directly resolves the two symmetry-related chiral sectors. In the CSR phase, spontaneous selection of a single Raman-cavity channel gives $\chi_c\simeq-1$ or $+1$, corresponding to $(\alpha_+,\alpha_-)\sim(\alpha,0)$ or $(0,\alpha)$, respectively. In contrast, the symmetric PW phases satisfy $|\alpha_+|^2=|\alpha_-|^2$ and hence $\chi_c=0$. A finite $\chi_c$ therefore directly signals spontaneous breaking of the channel-exchange symmetry, while its sign identifies the selected
propagation direction. Experimentally, $\chi_c$ can be extracted from the output intensities of the two counterpropagating cavity modes, providing a direct optical probe of the spontaneously selected chiral sector.

\begin{figure}
\centering
\includegraphics[width=0.95\columnwidth]{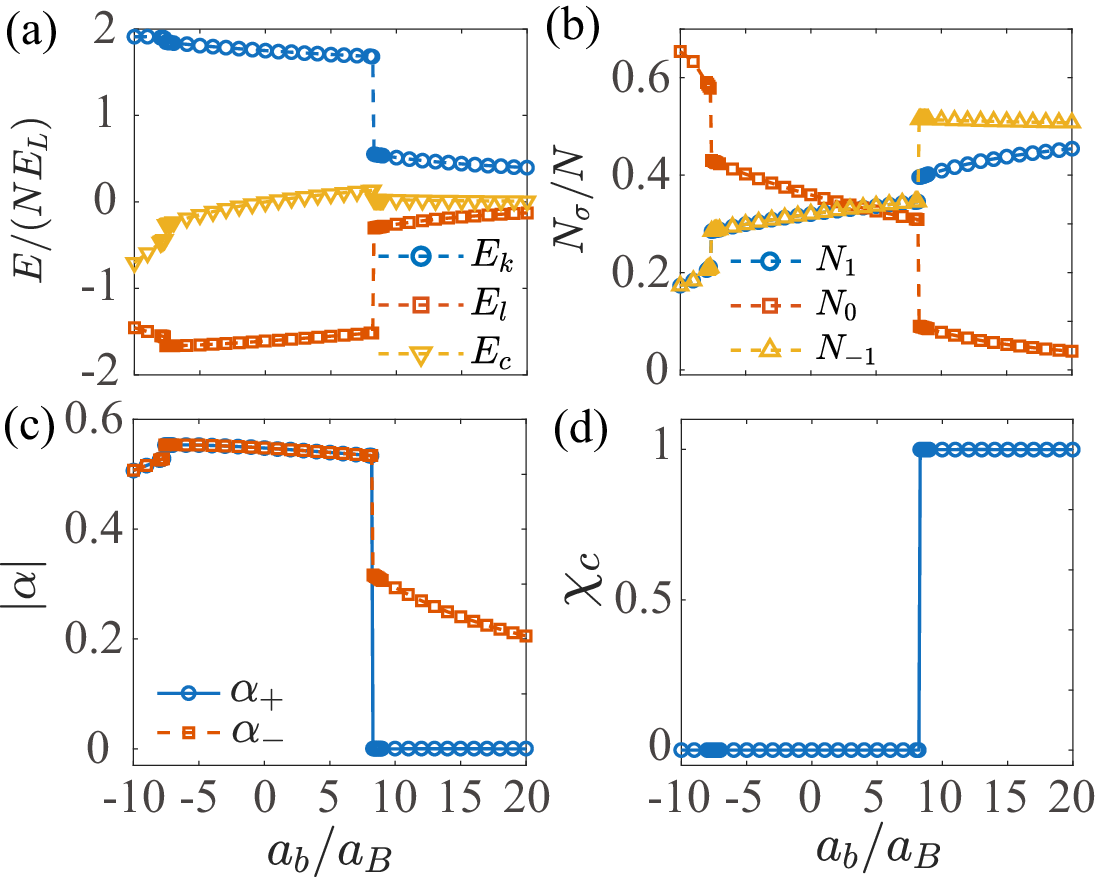}%
\caption{(a) Energy, (b) spin populations, (c) cavity-field amplitudes, and (d) chiral order parameter $\chi_c$ versus $a_b$ for $\hbar\Omega_0/E_L=5.5$.}
\label{gt}
\end{figure}

To elucidate the mechanism of spontaneous channel selection in the CSR phase, we analyze the effective energy landscape of the two competing Raman-cavity channels. As a diagnostic of the channel-selection instability, we introduce a small population imbalance $\epsilon$ between the two symmetry-related sectors when constructing the reduced energy functional. The parameter $\epsilon$ serves only as an infinitesimal symmetry-breaking seed and is not part of the physical Hamiltonian. Comparing the energies with and without this seed reveals whether the symmetric configuration is stable against an infinitesimal imbalance between the two channels.

Figure~\ref{pot}(a) compares the minimum effective potential for $\epsilon=0$ and $\epsilon=0.1N$ as functions of $\Omega_0$. Below the superradiant threshold, the two curves are nearly indistinguishable, indicating that the normal state is insensitive to the small channel imbalance. Upon approaching the superradiant regime, the imbalanced configuration becomes energetically favorable and reaches the instability at a lower value of $\Omega_0$. This behavior indicates an energetic tendency toward channel imbalance near the superradiant instability. The spontaneous nature of this selection is established more directly by the symmetry-related minima of the unperturbed energy landscape discussed below. The origin of this channel selection is seen more directly from the two-dimensional effective energy landscapes in Figs.~\ref{pot}(b) and \ref{pot}(c), shown for the CSR and PW2 phases, respectively.

For PW2, the interaction-induced inversion of the relative energetics of the Raman-coupled atomic configurations is incorporated into the corresponding effective energy functional. As derived in Appendix B, the interaction contribution relevant to the competition between the two symmetry-related channels contains the cross term $|\beta_1|^2|\beta_2|^2$, where $\beta_1=\langle\hat c_{0,\mathbf{k}_+}\rangle$ and $\beta_2=\langle\hat c_{0,\mathbf{k}_-}\rangle$ are the mean-field amplitudes of the two momentum modes associated with the two Raman-cavity channels. This term couples the two superradiant order parameters while preserving their exchange symmetry. In the CSR regime [Fig.~\ref{pot}(b)], the four degenerate minima lie on the coordinate axes of the $(\beta_1,\beta_2)$ plane. Each minimum therefore has either $\beta_1=0$ or $\beta_2=0$, corresponding to condensation into only one of the two symmetry-related Raman-cavity channels. Because the energy landscape itself remains invariant under $\beta_1\leftrightarrow\beta_2$, selection of one axis constitutes spontaneous breaking of the channel-exchange symmetry. The two minima on each axis differ only by the sign of the corresponding real order parameter, whereas the choice between the two axes distinguishes the two chiral sectors. In the PW2 regime [Fig.~\ref{pot}(c)], by contrast, the minima occur at finite $|\beta_1|$ and $|\beta_2|$ with equal magnitudes, corresponding to coexistence of the two Raman-cavity channels. The channel-exchange symmetry is therefore preserved, while the internal-state occupation is inverted relative to PW1.

Figures~\ref{pot}(d) and \ref{pot}(e) show angular cuts of the corresponding landscapes along the dashed circles in panels (b) and (c). The CSR minima occur along the coordinate axes, whereas the PW2 minima lie along directions with equal contributions from the two channels, making the distinct symmetry structures explicit. Consistent with these landscapes, the two channel-associated spin populations are equal in PW2 but become unequal in CSR, as independently seen in Figs.~\ref{pd}(c) and \ref{gt}(b).

\begin{figure}
\includegraphics[width=0.95\columnwidth]{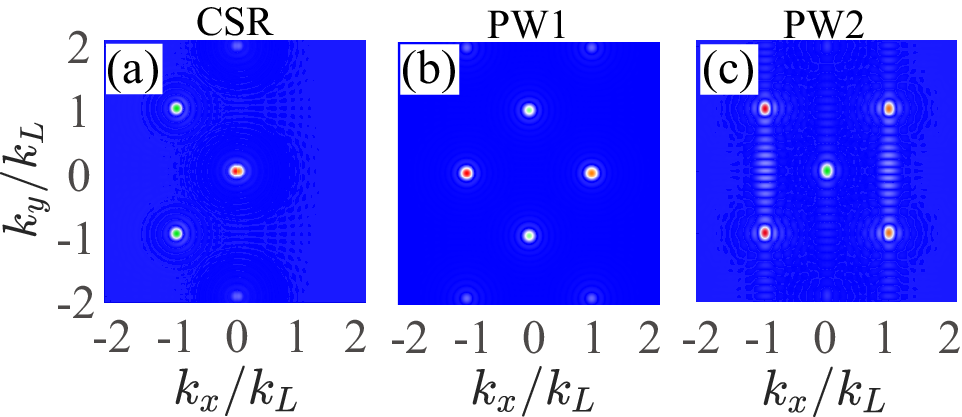}%
\caption{(a)-(c) Spin-resolved momentum distributions in the CSR, PW1, and PW2 phases, respectively, for $(\hbar\Omega_0/E_L,a_b/a_B)=(5.4,10)$, $(5.6,10)$, and $(4.5,-10)$. Green, red, and orange distributions denote the $m_F=0,+1$, and $-1$ components, respectively.}
\label{mm}
\end{figure}

We next examine the interaction-driven evolution at fixed Raman coupling. Figure~\ref{gt} shows the energy, atomic populations, cavity photon numbers, and chiral order parameter as functions of $a_b$. The evolution is governed primarily by the interaction-induced renormalization of the relative energies of the Raman-coupled spin configurations, or equivalently by the effective detuning $\widetilde{\delta}(g_b)$ introduced above. As $g_b$ is varied, the resulting change in $\widetilde{\delta}$ continuously redistributes the spin populations until a phase boundary is reached, where the $m_F=0$ population changes discontinuously [Fig.~\ref{gt}(b)]. The same interaction-induced level renormalization changes the relative stability of the CSR and PW1 states and thereby drives the CSR-PW1 transition at fixed $\Omega_0$. 

As shown in Fig.~\ref{gt}(d), $|\chi_c|$ becomes close to unity in the CSR regime and vanishes in the symmetric PW phases. The CSR-PW1 transition exhibits the same qualitative signatures whether it is crossed by varying $\Omega_0$ or $a_b$: discontinuous changes occur in the atomic populations and cavity occupations. For sufficiently negative $a_b$, the interaction-induced inversion of the internal-state energetics instead stabilizes the PW2 phase. Like PW1, PW2 preserves the channel-exchange symmetry and therefore has $\chi_c=0$; the distinction between the two phases lies instead in the reversed internal-state and momentum configuration of the Raman process, as illustrated in Fig.~\ref{model}(c).

\begin{figure}
\includegraphics[width=0.95\columnwidth]{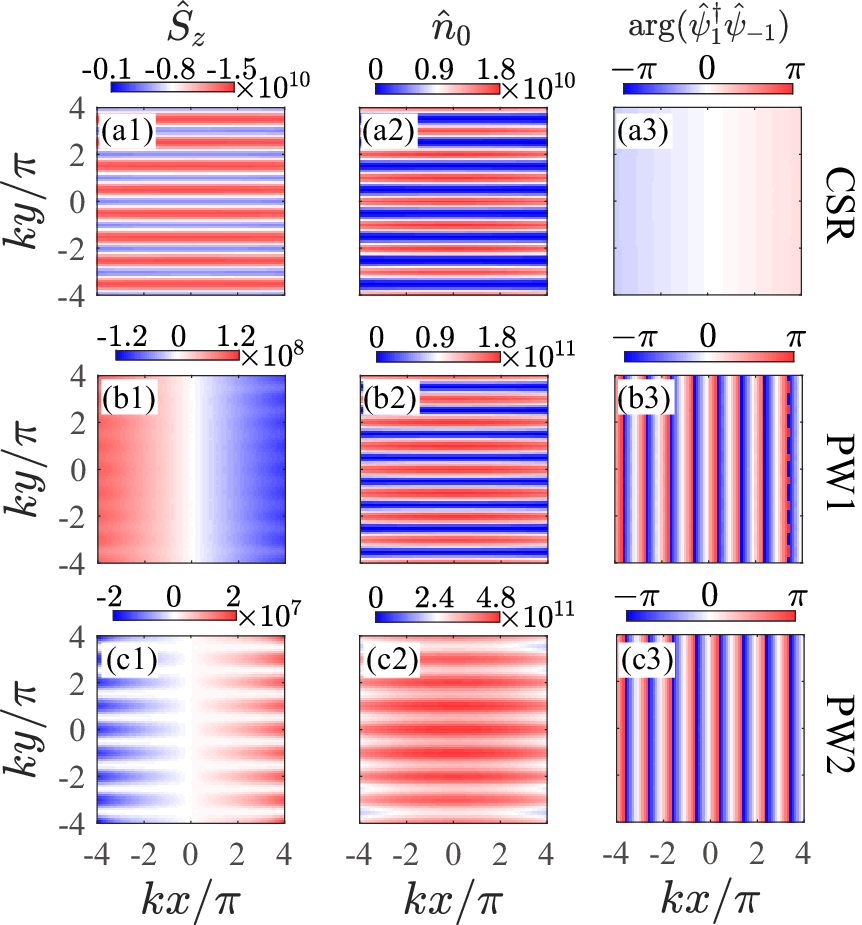}%
\caption{Columns 1-3 show the density difference between the $m_F=+1$ and $-1$ components, the $m_F=0$ density, and the relative phase between the $m_F=-1$ and $+1$ components, respectively. Rows 1-3 correspond to the CSR, PW1, and PW2 phases, respectively, for $(\hbar\Omega_0/E_L,a_b/a_B)=(5.4,10)$, $(5.6,10)$, and $(4.5,-10)$.}
\label{SS}
\end{figure}

Figures~\ref{mm} and \ref{SS} further characterize the CSR, PW1, and PW2 phases through their spin-resolved momentum distributions, densities, and relative phases. A key feature of the spin-$1$ system is that spin-dependent interactions constrain the relative spin populations and thereby reshape the superradiant phase structure. The spin-independent interaction $g_s$ primarily affects the overall density distribution and has only a weak influence on the phase boundaries considered here. The $g_a S_z^2$ interaction, by contrast, directly penalizes or favors an imbalance between the $m_F=\pm1$ components. For $g_a>0$, as considered here, the interaction energy is minimized by nearly equal populations of these two components. Accordingly, the N phase has approximately balanced $m_F=\pm1$ populations. For $g_a<0$, the same interaction instead favors spin polarization and hence an imbalance between them. For the positive $g_a$ considered here, there is therefore an energetic tendency toward balanced $m_F=\pm1$ populations, which competes with the spontaneous single-channel selection in the CSR phase.

In the PW1 and PW2 phases, both Raman-cavity channels are occupied and the channel-exchange symmetry is preserved. This symmetry is evident in the momentum distributions [Figs.~\ref{mm}(b) and \ref{mm}(c)] and densities [Figs.~\ref{SS}(b1) and \ref{SS}(c1)], where the $m_F=\pm1$ components remain nearly balanced. The residual density difference is two to three orders of magnitude smaller than the individual spin densities. In PW1 [Fig.~\ref{mm}(b)], several recoil-momentum components are appreciably occupied, so that the single-recoil truncation underlying the reduced model is no longer quantitatively adequate. In PW2 [Fig.~\ref{mm}(c)], by contrast, the dominant $m_F=0$ component is centered at zero momentum, while the weakly populated $m_F=\pm1$ components occupy the corresponding single-recoil modes, consistent with the reduced PW2 model. Their weak populations also lead to a relatively weak density modulation [Fig.~\ref{SS}(c2)].

The CSR phase exhibits a qualitatively different momentum structure [Fig.~\ref{mm}(a)]. The $m_F=\pm1$ components are concentrated near $(k_x,k_y)=(0,0)$, whereas the $m_F=0$ component occupies only the recoil branch $(-k_L,\pm k_L)$ associated with the selected Raman-cavity channel. The absence of the symmetry-related recoil branch is therefore a direct momentum-space signature of single-channel superradiance. The two $k_y=\pm k_L$ recoil components further produce a $\lambda/2$-periodic density modulation along $y$ [Fig.~\ref{SS}(a2)].

The relative-phase distributions provide an independent signature of the same channel selection. For the CSR branch shown in Fig.~\ref{SS}(a3), the occupied channel satisfies $\arg\!\left[\psi_0^*(\mathbf r)\psi_{+1}(\mathbf r)\alpha_+ e^{ik_Lx}\right]=\pi$, which minimizes the corresponding Raman-coupling energy. The symmetry-related channel exhibits no corresponding phase locking because $\alpha_-\simeq0$. Momentum selection and phase locking thus provide complementary signatures of single-channel superradiance. By channel-exchange symmetry, the opposite configuration with $\alpha_+\simeq0$ and $\alpha_-\neq0$ is exactly degenerate.

With increasing $\Omega_0$, occupation of the second Raman channel eventually becomes energetically favorable, driving the first-order CSR-to-PW1 transition shown in Fig.~\ref{pd}(b). In the PW phases, both Raman channels are occupied and satisfy $\arg\!\left[\psi_0^*(\mathbf r)\psi_{\pm1}(\mathbf r)\alpha_\pm e^{\pm ik_Lx}\right]=\pi$, thereby minimizing the corresponding Raman-coupling energies. Because the two cavity modes carry opposite momenta, simultaneous occupation of both channels produces the spatial phase winding between the $m_F=\pm1$ components observed in Figs.~\ref{SS}(b3) and \ref{SS}(c3). Such a winding is strongly suppressed in CSR, where only one Raman channel is phase locked to a finite cavity field.

Finally, the PW2 momentum distribution provides a direct signature of the interaction-induced inversion of the internal-state hierarchy discussed above. Although the bare two-photon detuning favors the $m_F=\pm1$ manifold, sufficiently negative $g_b$ reverses the effective level ordering, making $m_F=0$ the dominant low-energy component. Accordingly, the $m_F=0$ component carries most of the population at zero momentum, while the weakly populated $m_F=\pm1$ components occupy the corresponding recoil modes [Fig.~\ref{mm}(c)]. This momentum structure is consistent with the two symmetry-related Tavis-Cummings channels of the reduced PW2 Hamiltonian and shows that spin-dependent collisions can reconstruct the internal-state character of the superradiant phase rather than merely shift its threshold.

\section{Conclusion}

In summary, we have investigated the superradiant phase structure of a spin-$1$ Bose-Einstein condensate coupled to a ring cavity. Above the superradiant threshold, we identify a chiral superradiant (CSR) phase in which the channel-exchange symmetry of the Hamiltonian is spontaneously broken. The two symmetry-related CSR configurations correspond to spontaneous selection of either Raman-cavity channel, with one counterpropagating cavity mode macroscopically occupied and the other remaining dark. The effective energy landscape reveals this mechanism through symmetry-related minima corresponding to single-channel occupation. Spin-dependent collisions play a complementary role: they leave the N-to-CSR threshold comparatively weakly affected while substantially reshaping the boundaries between the CSR and PW phases. For sufficiently negative $a_b$, the interaction-renormalized two-photon detuning can change sign, producing an inversion of the internal-state hierarchy and stabilizing the PW2 phase with a dominant zero-momentum $m_F=0$ component. Our results establish spontaneous selection between degenerate superradiant channels as a mechanism for generating chirality in cavity QED and identify spin-dependent interactions as a means of controlling the resulting many-body phase structure. The associated chirality can be directly inferred from the directional cavity outputs, providing an experimentally accessible probe of the broken channel symmetry.

\begin{acknowledgments}
This work was supported by the National Key Research and Development Program of China (Grants No. 2025ZD0300400), NSFC (Grants No. 12674338, No. 12474270, and No. 12574295),  Ningbo Science $\&$ Technology Bureau (Grants No. 2024DJRC010009), and Guangdong Provincial Quantum Science Strategic Initiative (GDZX2604004).
\end{acknowledgments}

\appendix   
\setcounter{table}{0}   %从0开始编号，显示出来表会A1开始编号
\setcounter{figure}{0}
\setcounter{equation}{0}
%定义编号格式，在数字序号前加字符“A"
% \renewcommand{\thetable}{A\arabic{table}}
\renewcommand{\thefigure}{A\arabic{figure}}

\section{Effective atom-cavity Hamiltonian}

In this Appendix, we derive the effective atom-cavity Hamiltonian for the laser configuration and atomic level scheme shown in Fig.~1 of the main text. Within the rotating-wave approximation and omitting the atomic kinetic and trapping terms at this stage, the atom-cavity Hamiltonian reads
\begin{align}
{\hat H}/{\hbar}={}&\omega_m\hat e^\dagger\hat e+\omega_c\sum_{\nu=\pm}\hat a_\nu^\dagger\hat a_\nu+\delta\,\hat b_0^\dagger\hat b_0 \nonumber\\
&+\Omega\cos(k_Ly)\left(e^{-i\omega t}\hat e^\dagger\hat b_0+\mathrm{H.c.}\right) \nonumber\\
&+g\left[e^{ik_Lx}\hat e^\dagger\hat b_{+1}\hat a_++e^{-ik_Lx}\hat e^\dagger\hat b_{-1}\hat a_-+\mathrm{H.c.}\right],
\label{app:Hlab}
\end{align}
where $\hat a_\pm$ denote the annihilation operators of the two counterpropagating cavity modes, while $\hat b_\sigma$ and $\hat e$ annihilate an atom in the ground state $|\sigma\rangle$ ($\sigma=0,\pm1$) and the electronically excited state, respectively. The frequencies of the excited atomic state and the cavity modes are denoted by $\omega_m$ and $\omega_c$, respectively. The parameter $\delta$ denotes the effective two-photon detuning of the $m_F=0$ state relative to the $m_F=\pm1$ manifold, consistent with the convention used in the main text. The spatially dependent atom-cavity and pump couplings are $g_\pm(x)=g e^{\pm ik_Lx}$ and $\Omega(y)=\Omega\cos(k_Ly)$, respectively.

Transforming to the frame rotating at the pump frequency $\omega$ with $\hat U(t)=\exp\!\left[-i\omega t\left(\hat e^\dagger\hat e+\hat a_+^\dagger\hat a_++\hat a_-^\dagger\hat a_-\right)\right]$, gives
\begin{align}
{\hat H}/{\hbar}={}&\Delta\hat e^\dagger\hat e+\Delta_c\sum_{\nu=\pm}\hat a_\nu^\dagger\hat a_\nu+\delta\,\hat b_0^\dagger\hat b_0 \nonumber\\
&+\Omega\cos(k_Ly)\left(\hat e^\dagger\hat b_0+\mathrm{H.c.}\right) \nonumber\\
&+g\left[e^{ik_Lx}\hat e^\dagger\hat b_{+1}\hat a_++e^{-ik_Lx}\hat e^\dagger\hat b_{-1}\hat a_-+\mathrm{H.c.}\right],
\label{app:Hrot}
\end{align}
where $\Delta=\omega_m-\omega$ and $\Delta_c=\omega_c-\omega$ are the atom-pump and cavity-pump detunings, respectively.

Including the cavity decay rate $\kappa$ and the excited-state decay rate $\gamma$, the equations relevant to the elimination of the electronically excited state are
\begin{align}
i\dot{\hat e}={}&(\Delta-i\gamma)\hat e+\Omega\cos(k_Ly)\hat b_0 \nonumber \\
&+g\left(e^{ik_Lx}\hat a_+\hat b_{+1}+e^{-ik_Lx}\hat a_-\hat b_{-1}\right),
\label{app:eom_e} \nonumber\\
i\dot{\hat a}_+={}&(\Delta_c-i\kappa)\hat a_++g e^{-ik_Lx}\hat b_{+1}^\dagger\hat e,\nonumber\\
i\dot{\hat a}_-={}&(\Delta_c-i\kappa)\hat a_-+g e^{ik_Lx}\hat b_{-1}^\dagger\hat e.
\end{align}
In the far-detuned regime $|\Delta|\gg\{g,\Omega,\gamma\}$, the electronically excited state follows the ground-state and cavity dynamics adiabatically and can be eliminated by setting $\dot{\hat e}\simeq0$. This gives
\begin{equation}
\hat e\simeq-\frac{\Omega\cos(k_Ly)\hat b_0+g e^{ik_Lx}\hat a_+\hat b_{+1}+g e^{-ik_Lx}\hat a_-\hat b_{-1}}{\Delta-i\gamma}.
\label{app:ead}
\end{equation}
For $|\Delta|\gg\gamma$, we retain the leading dispersive contribution $1/(\Delta-i\gamma)\simeq1/\Delta$. Substituting Eq.~(\ref{app:ead}) back into the equations of motion yields the effective single-particle Hamiltonian given in Eq.~(\ref{single}) of the main text.

Including the atomic kinetic energy, trapping potential, and contact interactions, the many-body Hamiltonian can be written as $\hat{\mathcal H}=\hat{\mathcal H}_0+\hat{\mathcal H}_{\rm int}$, with
\begin{align}
\hat{\mathcal H}_0={}&\int d^3r\,\hat{\boldsymbol\Psi}^\dagger(\mathbf r)\left[-\frac{\hbar^2\nabla^2}{2M}+V_T(\mathbf r)+\hbar\hat{\boldsymbol h}(\mathbf r)\right]\hat{\boldsymbol\Psi}(\mathbf r) \nonumber\\
&+\hbar\Delta_c\left(\hat a_+^\dagger\hat a_++\hat a_-^\dagger\hat a_-\right),
\label{app:H0}
\end{align}
where $\hat{\boldsymbol\Psi}=(\hat\psi_{+1},\hat\psi_0,\hat\psi_{-1})^T$. The contact interaction takes the compact form
\begin{align}
\hat{\mathcal H}_{\rm int}={}&\frac{g_s}{2}\int d^3r\,\hat n^2(\mathbf r) +\frac{g_b}{2}\int d^3r\,\hat n_0^2(\mathbf r) \nonumber\\
&+\frac{g_a}{2}\int d^3r\,\!\left[\hat n_{+1}(\mathbf r)-\hat n_{-1}(\mathbf r)\right]^2\!,
\label{app:Hint}
\end{align}
where $\hat n_\sigma(\mathbf r)=\hat\psi_\sigma^\dagger(\mathbf r)\hat\psi_\sigma(\mathbf r)$ and $\hat n(\mathbf r)=\sum_\sigma\hat n_\sigma(\mathbf r)$. The effective single-particle coupling matrix is
\begin{equation}
\hat{\boldsymbol h}(\mathbf r)=
\begin{pmatrix}
M_0(y) & M_+^\dagger & U_0\hat a_+^\dagger\hat a_-e^{-2ik_Lx} \\
M_+ & M_0(y)+\delta & M_- \\
U_0\hat a_-^\dagger\hat a_+e^{2ik_Lx} & M_-^\dagger & M_0(y)
\end{pmatrix},
\label{app:hsp}
\end{equation}
where
\begin{align}
M_0(y)&=U_0\left(\hat a_+^\dagger\hat a_++\hat a_-^\dagger\hat a_-\right)+U_y\cos^2(k_Ly), \nonumber \\
M_\pm&=\Omega_0\cos(k_Ly)e^{\pm ik_Lx}\hat a_\pm. \nonumber 
\label{app:Mpm}
\end{align}
Here $U_0=-g^2/\Delta$ and $U_y=-\Omega^2/\Delta$ are the cavity- and pump-induced ac Stark shifts, respectively, while $\Omega_0=-g\Omega/\Delta$ is the effective two-photon Raman coupling strength.

At the mean-field level, we write $\psi_\sigma(\mathbf r)=\langle\hat\psi_\sigma(\mathbf r)\rangle$ and define the atomic overlap integrals
\begin{align}
\mathcal{G}
&=\int d^3r\,e^{-2ik_Lx}\psi_{+1}^*(\mathbf r)\psi_{-1}(\mathbf r), \nonumber\\
\Xi_{+1}
&=\int d^3r\,\cos(k_Ly)e^{-ik_Lx}\psi_{+1}^*(\mathbf r)\psi_0(\mathbf r), \nonumber\\
\Xi_{-1}
&=\int d^3r\,\cos(k_Ly)e^{ik_Lx}\psi_{-1}^*(\mathbf r)\psi_0(\mathbf r).
\end{align}
The cavity-field equations of motion then take the form
\begin{align}
i\dot{\alpha}_+
&=(\widetilde{\Delta}_c-i\kappa)\alpha_+
+U_0\mathcal{G}\,\alpha_-
+\Omega_0\Xi_{+1}, \nonumber\\
i\dot{\alpha}_-
&=(\widetilde{\Delta}_c-i\kappa)\alpha_-
+U_0\mathcal{G}^*\,\alpha_+
+\Omega_0\Xi_{-1}.
\label{app:cavity_eom}
\end{align}
where $\alpha_\pm=\langle\hat a_\pm\rangle$ and $\widetilde{\Delta}_c=\Delta_c+U_0N$ is the dispersively shifted cavity detuning.

In the fast-cavity regime, where the cavity dynamics is much faster than the external atomic motion, the cavity fields can be adiabatically eliminated~\cite{RevModPhys.85.553,Mivehvar:2021lpo}. Setting $\dot{\alpha}_\pm=0$ gives the steady-state coherent cavity amplitudes
\begin{align}
\alpha_+
&=\frac{\Omega_0\left[U_0\mathcal{G}\,\Xi_{-1}-(\widetilde{\Delta}_c-i\kappa)\Xi_{+1}\right]}{D_c}, \nonumber\\
\alpha_-
&=\frac{\Omega_0\left[U_0\mathcal{G}^*\,\Xi_{+1}-(\widetilde{\Delta}_c-i\kappa)\Xi_{-1}\right]}{D_c}.
\label{app:alpha}
\end{align}
where $D_c=(\widetilde{\Delta}_c-i\kappa)^2-U_0^2|\mathcal{G}|^2$. The corresponding intracavity photon numbers are $n_\pm=|\alpha_\pm|^2$.

\section{Effective quantum-optical model}

To elucidate the microscopic origin of the channel selection, we reduce the many-body Hamiltonian to an effective quantum-optical model within the single-recoil approximation~\cite{baumann2010dicke}. For the CSR phase, we retain the zero-momentum modes of the $m_F=\pm1$ components and the lowest recoil modes of the $m_F=0$ component. For a homogeneous condensate of volume $V$, the atomic field operators are expanded as
\begin{align}
\hat{\psi}_0(\mathbf r)
&=\sqrt{\frac{2}{V}}\cos(k_Ly)
\left(
e^{ik_Lx}\hat c_{0,\mathbf k_+}
+e^{-ik_Lx}\hat c_{0,\mathbf k_-}
\right),
\nonumber\\
\hat{\psi}_{+1}(\mathbf r)
&=\frac{1}{\sqrt V}\hat c_{+,\mathbf0},
\qquad
\hat{\psi}_{-1}(\mathbf r)
=\frac{1}{\sqrt V}\hat c_{-,\mathbf0},
\label{app:csr_expansion}
\end{align}
where $V$ is the condensate volume, $\hat c_{\sigma,\mathbf k}$ annihilates an atom in spin state $\sigma$ and momentum mode $\mathbf k$, while $\mathbf k_\pm$ label the recoil modes associated with the two Raman-cavity channels

Neglecting the trapping potential and collisional interactions, substitution of Eq.~(\ref{app:csr_expansion}) into the many-body Hamiltonian yields two symmetry-related Tavis--Cummings-type channels,
\begin{align}
{\hat H_{\rm CSR}}/{\hbar}
={}&\widetilde{\Delta}_c
\sum_{\nu=\pm}\hat a_\nu^\dagger\hat a_\nu
\nonumber\\
&+\frac{\omega_0}{2}
\sum_{\nu=\pm}
\left(
\hat c_{0,\mathbf k_\nu}^\dagger
\hat c_{0,\mathbf k_\nu}
-\hat c_{\nu,\mathbf0}^\dagger
\hat c_{\nu,\mathbf0}
\right)
\nonumber\\
&+\frac{\Omega_0}{\sqrt2}
\sum_{\nu=\pm}
\left(
\hat a_\nu
\hat c_{0,\mathbf k_\nu}^\dagger
\hat c_{\nu,\mathbf0}
+\mathrm{H.c.}
\right),
\label{app:Hcsr}
\end{align}
where $\omega_0=2E_L/\hbar+\delta$ is the effective atomic detuning. Equation~(\ref{app:Hcsr}) explicitly displays two symmetry-related Raman-cavity channels with the same coupling strength $\Omega_0/\sqrt2$.

At the mean-field level, we define $\beta_1=\langle\hat c_{0,\mathbf k_+}\rangle$ and $\beta_2=\langle\hat c_{0,\mathbf k_-}\rangle$. After adiabatic elimination of the cavity fields, the corresponding effective potential takes the form
\begin{align}
V_g^{\rm CSR}
={}&\hbar\omega_0
\left(
|\beta_1|^2+|\beta_2|^2
\right)
\nonumber\\
&-\frac{\hbar\Omega_0^2}{2\widetilde{\Delta}_c}
|\beta_1|^2
\left(
N_{+1}-|\beta_1|^2
\right)
\nonumber\\
&-\frac{\hbar\Omega_0^2}{2\widetilde{\Delta}_c}
|\beta_2|^2
\left(
N_{-1}-|\beta_2|^2
\right)
+g_e|\beta_1|^2|\beta_2|^2,
\label{app:Vcsr}
\end{align}
where $N_{\pm1}$ denote the populations of the $m_F=\pm1$ components. To probe the stability of the symmetric configuration against channel imbalance, we introduce a population perturbation $N_{\pm1}=\frac{N}{2}\pm\epsilon$. The coefficient $g_e$ describes the interaction contribution that directly couples the occupations of the two Raman-cavity channels.

For compactness, we introduce the mode occupations $\hat n_\pm =\hat c_{\pm,\mathbf0}^\dagger\hat c_{\pm,\mathbf0}$ and $\hat n_{0,\pm}
=\hat c_{0,\mathbf k_\pm}^\dagger \hat c_{0,\mathbf k_\pm}$. Within the single-recoil expansion, the $g_a$ contribution becomes
\begin{align}
&\frac{g_a}{2}
\int d^3r\,
(\hat n_{+1}-\hat n_{-1})^2
={}
\frac{g_a}{2V}
\left(
\hat n_+^2+\hat n_-^2
-2\hat n_+\hat n_-
\right)
\nonumber\\
={}&
\frac{g_a}{2V}
\left\{
\left(
\frac{N}{2}-|\beta_1|^2
\right)^2
+
\left(
\frac{N}{2}-|\beta_2|^2
\right)^2
\right.
\nonumber\\
&\left.
-2
\left[
\frac{N^2}{4}
-\frac{N}{2}
\left(
|\beta_1|^2+|\beta_2|^2
\right)
+|\beta_1|^2|\beta_2|^2
\right]
\right\},
\label{app:ga_reduced}
\end{align}
Similarly, the $g_b$ contribution is
\begin{align}
\frac{g_b}{2}
\int d^3r\,\hat n_0^2
={}&
\frac{3g_b}{4V}
\left[
(\hat n_{0,+}+\hat n_{0,-})^2
+2\hat n_{0,+}\hat n_{0,-}
\right]
\nonumber\\
={}&
\frac{3g_b}{4V}
\left(
|\beta_1|^4+|\beta_2|^4
+4|\beta_1|^2|\beta_2|^2
\right),
\label{app:gb_reduced}
\end{align}
Collecting the terms proportional to $|\beta_1|^2|\beta_2|^2$ gives $g_e=({3g_b-g_a})/{V}$. The remaining interaction terms depend separately on $|\beta_1|^2$ or $|\beta_2|^2$ and therefore do not directly couple the two channel occupations. To isolate the interaction contribution responsible for channel competition, we retain the cross term $g_e|\beta_1|^2|\beta_2|^2$ in Eq.~(\ref{app:Vcsr}). This reduced potential provides a qualitative description of the single-channel selection underlying the CSR phase.

For comparison, we perform an analogous single-recoil expansion for the PW2 phase,
\begin{align}
\hat{\psi}_0(\mathbf r)
&=\frac{1}{\sqrt V}\hat d_0,
\nonumber\\
\hat{\psi}_{+1}(\mathbf r)
&=\sqrt{\frac{2}{V}}
\cos(k_Ly)e^{-ik_Lx}\hat d_+,
\nonumber\\
\hat{\psi}_{-1}(\mathbf r)
&=\sqrt{\frac{2}{V}}
\cos(k_Ly)e^{ik_Lx}\hat d_-,
\label{app:pw2_expansion}
\end{align}
where $\hat d_0$ annihilates an atom in the zero-momentum $m_F=0$ mode, while $\hat d_\pm$ annihilate atoms in the recoil modes of the $m_F=\pm1$ components, respectively. The corresponding reduced Hamiltonian is
\begin{align}
{\hat H_{\rm PW2}}/{\hbar}
={}&
\widetilde{\Delta}_c
\sum_{\nu=\pm}
\hat a_\nu^\dagger\hat a_\nu
\nonumber\\
&+\frac{\omega_0}{2}
\left(
\hat d_+^\dagger\hat d_+
+\hat d_-^\dagger\hat d_-
-\hat d_0^\dagger\hat d_0
\right)
\nonumber\\
&+\frac{\Omega_0}{\sqrt2}
\left(
\hat a_+\hat d_0^\dagger\hat d_+
+\hat a_-\hat d_0^\dagger\hat d_-
+\mathrm{H.c.}
\right)
\nonumber\\
&+U_0
\left(
\hat a_+^\dagger\hat a_-
\hat d_+^\dagger\hat d_-
+\mathrm{H.c.}
\right).
\label{app:Hpw2}
\end{align}

At the mean-field level, we define $\beta_1=\langle\hat d_+\rangle$ and $\beta_2=\langle\hat d_-\rangle$. Adiabatic elimination of the cavity fields gives the effective potential
\begin{align}
V_g^{\rm PW2}
={}&
\frac{
\hbar U_0\Omega_0^2 \left(
N-|\beta_1|^2-|\beta_2|^2
\right)
|\beta_1|^2|\beta_2|^2
}{
D_{\rm PW2}
}
\nonumber\\
&-
\frac{
\hbar\widetilde{\Delta}_c\Omega_0^2 \left(
N-|\beta_1|^2-|\beta_2|^2
\right)
\left(
|\beta_1|^2+|\beta_2|^2
\right)
}{
2D_{\rm PW2}
}
\nonumber\\
&+
\hbar\omega_0
\left(
|\beta_1|^2+|\beta_2|^2
\right),
\label{app:Vpw2}
\end{align}
with $D_{\rm PW2}=\widetilde{\Delta}_c^2-U_0^2|\beta_1|^2|\beta_2|^2$.

%\bibliography{nonreci_sr}

\end{document}